\documentclass[aps,twocolumn,pra,groupedaddress,superscriptaddress,nofootinbib]{revtex4}

\usepackage{graphicx,amsmath,amssymb,amsbsy,subfigure,hyperref,bbm,times,txfonts,float}
\usepackage{subfigure,hyperref,bbm,times}
\usepackage[T1]{fontenc}
\usepackage{braket}
\usepackage{epsfig}
\usepackage{color}
\usepackage{graphicx}
\usepackage{dcolumn}
\usepackage{bm}

\providecommand{\openone}{\leavevmode\hbox{\small1\kern-3.8pt\normalsize1}}

\usepackage{soul}

\hypersetup{
   colorlinks=true,
   linkcolor=blue,
}
\graphicspath{{figure/}}

\begin{document}


\title{Witnessing non-Markovian effects of quantum processes through Hilbert-Schmidt speed}

\author{Hossein Rangani Jahromi}
\email{h.ranganijahromi@jahromu.ac.ir}%
\affiliation{Physics Department, Faculty of Sciences, Jahrom University, P.B. 74135111, Jahrom, Iran}

\author{Kobra Mahdavipour}%
\affiliation{Dipartimento di Ingegneria, Universit\`{a} di Palermo, Viale delle Scienze, Edificio 9, 90128 Palermo, Italy}%
\affiliation{INRS-EMT, 1650 Boulevard Lionel-Boulet, Varennes, Qu\'{e}bec J3X 1S2, Canada}
\author{Mahshid Khazaei Shadfar}%
\affiliation{Dipartimento di Ingegneria, Universit\`{a} di Palermo, Viale delle Scienze, Edificio 9, 90128 Palermo, Italy}%
\affiliation{INRS-EMT, 1650 Boulevard Lionel-Boulet, Varennes, Qu\'{e}bec J3X 1S2, Canada}

\author{Rosario Lo Franco}%
 \email{rosario.lofranco@unipa.it}
\affiliation{Dipartimento di Ingegneria, Universit\`{a} di Palermo, Viale delle Scienze, Edificio 6, 90128 Palermo, Italy}%

\date{\today}

\begin{abstract}
Non-Markovian effects can speed up the dynamics of quantum systems while the limits of the evolution time can be derived by quantifiers of quantum statistical speed. We introduce a witness for characterizing the non-Markovianity of quantum evolutions through the Hilbert-Schmidt speed (HSS), which is a special type of quantum statistical speed. This witness has the advantage of not requiring diagonalization of evolved density matrix. Its sensitivity is investigated by considering several paradigmatic instances of open quantum systems, such as one qubit subject to phase-covariant noise and Pauli channel, two independent qubits locally interacting with leaky cavities, V-type and $\Lambda $-type three-level atom (qutrit) in a dissipative cavity. We show that the proposed HSS-based non-Markovianity witness detects memory effects in agreement with the well-established trace distance-based witness, being sensitive to system-environment information backflows.
\end{abstract}

       

\maketitle

\section{Introduction\label{introduction}}

The interaction of quantum systems with the surrounding environment leads to dissipating energy and losing quantum coherence \cite{breuer2002theory}.
Nevertheless, the process does not need to be monotonic and the quantum system may recover temporarily some of the lost energy or information due to memory effects during the evolution \cite{NM1,breuer2016colloquium,rivas2014quantum,lofrancoreview,mortezapourOpt,
RevModPhys.86.1203,GholipourAnnPhys,darrigo2012AOP,LoFrancoNatCom,
origin2,origin1,NMexp5,NMexp4,NMexp3,NMexp2,NMexp1}. This dynamical behavior, named non-Markovianity, can then act as a resource in various quantum information tasks such as teleportation with mixed states \cite{laine2014nonlocal}, improvement of capacity for long quantum channels \cite{bylicka2014non}, efficient entangling protocols \cite{xiang2014entanglement, mirkin1, mirkin2}, and work extraction from an Otto cycle \cite{thomas2018thermodynamics}. 

Characterization and quantification of non-Markovianity has been a subject of intense study \cite{rivas2014quantum,breuer2016colloquium,teittinen2018revealing,naikoo2019facets}. One route is to investigate temporary increases of the
entanglement shared by the open quantum system with an isolated ancilla,
which amounts to measure the deviation from complete positivity (CP-divisibility) of the dynamical map describing the evolution of the system \cite{rivas2010entanglement}. Another approach
\cite{breuer2009measure,laine2010measure} relies on measuring the distinguishability of two optimal initial states evolving through the same quantum channel and detecting any non-monotonicity (information backflows). 
Further witnesses of non-Markovianity have been proposed, based on different dynamical figures of merit, such as: negative time-dependent decoherence rates appearing in the canonical form of the master equation \cite{hall2014canonical}, channel capacities \cite{bylicka2014non}, quantum mutual information \cite{luo2012quantifying}, local quantum uncertainty \cite{he2014measuring}, quantum interferometric power \cite{dhar2015characterizing,girolami2014quantum,fanchini2017lectures,jahromi2019multiparameter}, coherence \cite{chanda2016delineating,he2017non}, state fidelity  \cite{jahromi2019multiparameter,rajagopal2010kraus,farajollahi2018estimation}, change of volume of the set of accessible states of the evolved system \cite{lorenzo2013geometrical}, Fisher information
flow \cite{lu2010quantum,rangani2017relation},  spectral analysis \cite{zhang2012general}, entropy production rates \cite{strasberg2019non,jahromi2015precision}, correlation measures \cite{de2019correlation}, Choi state \cite{zheng2020detecting} and quantum evolution speedup \cite{deffner2013quantum,xu2014quantum,xu2018hierarchical}.
This variety of witnesses and approaches highlight the multifaceted nature
of non-Markovian behavior which hence cannot be attributed to a unique feature of the system-environment interaction, preventing the characterization by means of a single tool for such a phenomenon.

CP-divisibility is the most common definition for Markovianity in open quantum systems \cite{breuer2002theory,rivas2014quantum}. A dynamical map $\{\mathcal{E}_{t}\}_{t\geq 0} $ is defined as a family of completely positive (CP) and trace-preserving (TP) maps acting on the system Hilbert space $ \mathcal{H} $. Generally speaking, one calls a map $k$-positive if the composite map $\mathcal{E}_{t}\otimes \mathbb{I}_{k}  $ is positive, where $ k $, $ \mathbb{I}_{k} $ denote the dimensionality of the
ancillary Hilbert space and its identity operator, respectively \cite{darek2019}. Provided that $\mathcal{E}_{t}\otimes \mathbb{I}_{k}  $ is positive for all $ k\geq 0 $ and for all $t$, then the dynamical map is completely positive. One then says that the dynamical map $ \mathcal{E}_{t} $ is CP-divisible (P-divisible) when the propagator $ V_{t,s} $, defined by $\mathcal{E}_{t}=V_{t,s}\circ \mathcal{E}_{s}  $, is completely positive (positive) for all $ t\geq s \geq 0 $ \cite{breuer2002theory}. According to the non-Markovianity measure introduced by Rivas-Huelga-Plenio (RHP) \cite{rivas2010entanglement}, the quantum evolution is considered Markovian if and only if the corresponding dynamical map $ \mathcal{E}_{t} $ is CP-divisible. 

The non-Markovian character of the system dynamics can be identified through another well-known perspective proposed by Breuer-Laine-Piilo (BLP), namely the distinguishability of two evolving quantum states of the same system \cite{breuer2009measure,laine2010measure}. This distinguishability is  quantified by the trace distance, a commonly used distance measure for two arbitrary states $ \rho_{1} $ and $ \rho_{2} $, defined as $D(\rho_{1},\rho_{2})=\frac{1}{2}\text{Tr}|\rho_{1}-\rho_{2}|$,
where $ |A|=\sqrt{A^{\dagger}A} $ for some operator $ A $. 
The trace distance $ D(\rho_{1},\rho_{2}) $ is contractive under
CPTP maps, i.e. $ D(\mathcal{E}_{t}(\rho_{1}),\mathcal{E}_{t}(\rho_{2}))\leq D(\rho_{1},\rho_{2})$.
Nevertheless, this does not mean generally that $ D(\mathcal{E}_{t}(\rho_{1}),\mathcal{E}_{t}(\rho_{2})) $ is a monotonically decreasing function of time. In fact, $\frac{\mathrm{d}}{\mathrm{d}t}D(\mathcal{E}_{t}(\rho_{1}),\mathcal{E}_{t}(\rho_{2})) > 0$ implies violation of P-divisibility and therefore of
CP-divisibility \cite{breuer2009measure,PhysRevA.90.022110}. In other words, under any Markovian evolution of the quantum system, one gets $ \mathrm{d} D(\mathcal{E}_{t}(\rho_{1}),\mathcal{E}_{t}(\rho_{2}))/{\mathrm{d}t}\leq 0$, owing to the contraction property. Therefore, its non-monotonicity can be understood as a witness of non-Markovianity due to system-environment backflows of information.

Studies on the role of typical figures of merit for quantum metrology, based on quantum Fisher information metric, to witness non-Markovianity have been also reported \cite{lu2010quantum,adesso2017}. On the other hand, non-Markovian effects can speed up the quantum evolution of a system \cite{deffner2013quantum,PhysRevA.94.052125,PhysRevA.93.020105,Ahansaz2019,adesso2017,wuPRA,Zou_2020}. 
It is known that quantifiers of statistical speed in the system Hilbert space may be associated with measures adopted in quantum metrology to investigate the ultimate limit of precision in estimating a given physical quantity \cite{braunstein1994statistical}. 
The sensitivity of an initial quantum state to changes of the parameter (e.g., an unknown phase shift) of a dynamical evolution can be then determined by measures of quantum statistical speed \cite{gessner2018statistical}. A higher sensitivity implies higher precision in the estimation of the parameter of interest \cite{braunstein1994statistical,giovannetti2006quantum,giovannetti2011}. 
These arguments naturally motivate one to inquire whether measures of quantum statistical speed can conveniently quantify the non-Markovian character of the system dynamics, a problem which has remained unexplored.

Here, we address this issue introducing a method for witnessing and measuring non-Markovianity by means of the Hilbert-Schmidt speed (HSS) \cite{gessner2018statistical}, a type of quantum statistical speed which has the advantage of avoiding diagonalization of the evolved density matrix. We check the efficiency of the proposed HSS-based witness in several typical situations of open quantum systems made of qubits and qutrits. In particular, we consider: one qubit subject to phase-covariant noise \cite{lankinen2016complete}, especially the so-called eternal non-Markovianity model \cite{he2017non,chruscinski2013non,chruscinski2015non,hall2014canonical,teittinen2019there};
a single qubit undergoing the Pauli channel \cite{breuer2016colloquium,chruscinski2013non,song2017dynamics}; two independent qubits locally interacting with leaky cavities; V-type and $\Lambda $-type three-level atom (qutrit) in a dissipative cavity. We find that the HSS-based non-Markovianity witness identifies memory effects in total agreement with the trace distance-based BLP witness, thus detecting system-environment information backflows.

The paper is organized as follows. In Sec.~\ref{sec:HSS} we briefly review the definition of the Hilbert-Schmidt speed. In Sec.~\ref{Witness} we introduce the measure of quantum non-Markovianity via the HSS. Through various examples, the sensitivity of this measure in detecting memory effects is studied in Sec.~\ref{Example}. 
Finally, Sec.~\ref{cunclusion} summarizes the main results and prospects.

\section{Hilbert-Schmidt speed (HSS)}\label{sec:HSS}

We start by recalling the general framework leading to the definition of quantum statistical speed, whose the HSS is a particular case. 

Let us consider the family of distance measures
\begin{equation}\label{dis}
[d_{\alpha}(p,q)]^{\alpha}=\dfrac{1}{2}\sum\limits_{x}^{}|p_{x}-q_{x}|^{\alpha},
\end{equation}
with $ \alpha\geq 1 $ and where $ p = \{p_{x}\}_{x} $ and $ q = \{q_{x}\}_{x} $ are probability distributions.
Here it is assumed that the random variable $ x $  takes only discrete values; 
in the case of a continuum of values, the sum is
replaced by an integral. These distances satisfy the following basic properties:
(i) non-negativity and normalization
$0 \leq d_{\alpha}(p,q)\leq 1 $, where $ d_{\alpha}(p,q)=0~\leftrightarrow p\equiv q $; (ii) triangle inequality $ d_{\alpha}(p_{1},p_{3})\leq d_{\alpha}(p_{1},p_{2})+d_{\alpha}(p_{2},p_{3}) $; (iii) symmetry $ d_{\alpha}(p,q)=d_{\alpha}(q,p) $.  

Generally, in order to obtain the statistical speed from any statistical distance, one should quantify the distance between infinitesimally close distributions taken from a one-parameter family $ p_{x}(\varphi) $ with parameter $ \varphi $. Then, the classical statistical speed is given by 
\begin{equation}\label{classicalspeed}
\text{s}_{\alpha}\big[p(\varphi_{0})\big]=\dfrac{\mathrm{d}}{\mathrm{d}\varphi}d_{\alpha}\big(p(\varphi_{0}+\varphi),p(\varphi_{0})\big).
\end{equation}

Considering now a given pair of quantum states $ \rho $ and $ \sigma $, one can extend these classical notions to the quantum case by taking 
$ p_{x} = \text{Tr}\{E_{x}\rho\} $ and $ q_{x} = \text{Tr}\{E_{x}\sigma\} $ as the measurement probabilities associated with the positive-operator-valued measure (POVM) defined by the set of $ \{E_{x}\geq 0\} $ satisfying $\sum_{x}^{} E_{x} = \mathbb{I}  $, where $\mathbb{I}$ is the identity operator. 
Maximizing the classical distance over all  possible choices of
POVMs, one obtains the corresponding quantum distance 
\begin{equation}\label{qdis}
D_{\alpha}(\rho,\sigma)=\max_{\{E_{x}\}}d_{\alpha}(\rho,\sigma),
\end{equation}
which leads to the expression \cite{gessner2018statistical}
\begin{equation}\label{qqdis}
[D_{\alpha}(\rho,\sigma)]^{\alpha}=\frac{1}{2}\text{Tr}|\rho-\sigma|^{\alpha},
\end{equation}
where $ |X|^{\alpha} $ can be computed using the spectral decomposition $ X\equiv\sum_{i}^{}\lambda_{i}|\lambda_{i} \rangle \langle  \lambda_{i}| $, i.e., $ |X|^{\alpha}=\sum_{i}^{}|\lambda_{i}|^{\alpha}|\lambda_{i} \rangle \langle  \lambda_{i}| $, so that $\text{Tr}|X|^{\alpha}=\sum_{i}^{}|\lambda_{i}|^{\alpha}  $.  For $ \alpha = 1 $, the trace distance $D(\rho_{1},\rho_{2})=\frac{1}{2}\text{Tr}|\rho_{1}-\rho_{2}|$ is retrieved, while for $ \alpha = 2 $ one gets the so-called Hilbert-Schmidt distance $D_{2}(\rho,\sigma)$ allowing for a simple evaluation because it does not need diagonalization of the argument operator. This distance is of Riemann type and limited by the following inequality relation
\begin{equation}
0 \leq  D_{2}(\rho,\sigma) \leq 2D(\rho,\sigma).
\end{equation}
The Hilbert-Schmidt distance generally does not possess the contractivity property, although quantum systems such as qubits constitute useful exceptions. Necessary and sufficient conditions for contractivity of the Hilbert-Schmidt distance for the Lindblad operators have been discussed \cite{wang2009contractivity}. For a single qubit, it is straightforward to derive that trace and Hilbert-Schmidt distances are equivalent, namely
\begin{equation}
D_{2}(\rho,\sigma)=\sqrt{2} D(\rho,\sigma),
\end{equation}
so that contractivity of trace distance implies contractivity of Hilbert-Schmidt distance. However, it worth to notice that this argument cannot be generalized to high-dimensional systems with Hilbert space dimension larger than two \cite{wang2009contractivity}.

Extending Eq.~(\ref{classicalspeed}) to the quantum case, one then obtains the quantum statistical speed as \cite{gessner2018statistical}
\begin{equation}\label{quantumspeed}
\text{S}_{\alpha}\big[\rho(\varphi)\big]=\max_{\{E_{x}\}} \text{s}_{\alpha}\big[p(\varphi)\big]=\bigg(\frac{1}{2}\text{Tr}\bigg|\dfrac{d\rho(\varphi)}{d\varphi}\bigg|^{\alpha}\bigg)^{1/\alpha}.
\end{equation}
In the special case when $ \alpha = 2 $, the quantum statistical speed is given by the Hilbert-Schmidt speed (HSS) \cite{gessner2018statistical} 
\begin{equation}\label{HSS}
HSS(\rho_{\varphi})=\sqrt{\frac{1}{2}\text{Tr}\bigg[\bigg(\dfrac{\text{d}\rho_{\varphi}}{\text{d}\varphi}\bigg)^2\bigg]},
\end{equation}
which, in analogy with the Hilbert-Schmidt distance, does not require the diagonalization of $ \text{d}\rho_{\varphi}/\text{d}\varphi $. 
Notice that noncontractivity of the Hilbert-Schmidt distance does not consequently imply noncontractivity of the HSS. In fact, on the one hand, the Hilbert-Schmidt distance is computed by maximization over all the possible choices of POVMs $\{E_x\}$ of the adopted distance measure $d_2(\rho,\sigma)$ (see Eqs.~(\ref{dis}) and (\ref{qdis})); on the other hand, the HSS is determined by maximization applied after the differentiation with respect to $\varphi$, starting from the adopted distance measure (see Eqs.~(\ref{classicalspeed}) and (\ref{quantumspeed})). Because of these computational subtleties, from the noncontractivity of the Hilbert-Schmidt distance, one cannot conclude that the HSS is also noncontractive. 
Indeed, we shall show in the following that the HSS can be regarded as a trustful, convenient non-Markovianity measure just because of its contractivity.

\section{HSS-based non-Markovianity measure \label{Witness}}

It is known that non-Markovian effects can lead to faster quantum evolution from an initial state to
a subsequent one \cite{deffner2013quantum,PhysRevA.94.052125,PhysRevA.93.020105,Ahansaz2019,wuPRA,Zou_2020}.
It thus seems natural that measures of quantum speed limits may play the role of proper quantifiers of memory effects occurring during a system dynamics. Some works along this direction based on quantum Fisher information metric have been reported \cite{lu2010quantum,adesso2017}. Here we aim at exploiting a convenient quantum statistical speed \cite{gessner2018statistical} as a figure of merit of the non-Markovian character of quantum evolutions, which avoids diagonalization of the system density matrix, with consequent practical advantages in the analysis. We stress that such a quantifier would be particularly useful, especially for detecting the memory effects of high-dimensional and multipartite open quantum systems. Looking at the various possible choices among the quantum statistical speeds of Eq.~(\ref{quantumspeed}), the most natural candidate towards this aim is just that obtained for 
$\alpha=2$, corresponding to the Hilbert-Schmidt speed (HSS) of Eq.~(\ref{HSS}). To assume the role of a faithful indicator of non-Markovianity, the HSS should not exhibit the problems of contractivity manifested by the Hilbert-Schmidt distance for dimensions larger than two \cite{wang2009contractivity}. We shall see that, interestingly, the HSS is indeed contractive at least for quantum systems having dimension $n\leq 3$.

In this regard, for a quantum system with $n$-dimensional Hilbert space $\mathcal{H}$, let us take an initial state defined as 
\begin{equation}\label{initialstate}
|\psi_{0}\rangle=\dfrac{1}{\sqrt{n}}\big(\text{e}^{i\varphi}|\psi_{1}\rangle+\ldots+|\psi_{n}\rangle\big),
\end{equation}
where $\varphi$ is an unknown phase shift and $\{|\psi_{i}\rangle,\ i=1,\ldots,n\}$ constructs a complete and orthonormal set (basis) for $ \mathcal{H} $. 
The form of $|\psi_{0}\rangle$ is strategically chosen for phase-sensitive quantum statistical speed, being the standard initial state structure for quantum metrology phase estimation \cite{giovannetti2006quantum,giovannetti2011}. 
With the idea that a nonmonotonic speed (positive acceleration) of the quantum dynamics is a signature of memory effects in the system dynamics, we then introduce the HSS-based witness of non-Markovianity as
\begin{equation}\label{chit}
\chi(t):= \dfrac{\text{d}HSS \big(\rho_{\varphi}(t)\big)}{\text{d}t} > 0,
\end{equation}
where $ \rho_{\varphi}(t) $ denotes the evolved state of the system and $HSS (\rho_{\varphi}(t))$ is defined in Eq.~(\ref{HSS}). 
Given this witness, in analogy to what has been done for other measures \cite{breuer2009measure,laine2010measure}, a quantifier of the degree of non-Markovianity can be naturally defined as
\begin{equation}\label{NHSS}
\mathcal{N}_\mathrm{HSS}:=\max_{{\varphi,\{|\psi_{1}\rangle,...,|\psi_{n}\rangle\}}} \int\limits_{\chi(t)>0}^{}\chi(t)\text{dt},
\end{equation}
where the maximization is taken over all the possible parametrizations of the single initial state of Eq.~(\ref{initialstate}). 

Notice that here we are interested in only detecting non-Markovian effects by the HSS-based witness, so that its actual value is not important and no optimization over the initial state parameters is required. 
The sanity check of $\chi(t)$ as faithful witness of non-Markovianity is performed in the following section. 

\section{Qualitative analysis of non-Markovianity}\label{Example}

In this section, we consider several typical examples of open quantum systems of both theoretical and experimental interest to qualitatively analyze the faithfulness of the HSS-based non-Markovianity witness defined above.
Notice that, to this aim, it is sufficient to verify that the HSS is contractive for memoryless dynamics and  sensitive to system-environment information backflows, occurring in correspondence of $\chi(t)>0$ (speedup of the quantum evolution, as identified by Eq.~(\ref{chit})). 
We shall study the time behavior of $\chi(t)$, verifying that whenever it is positive then the BLP (trace distance-based) witness $\sigma(t)\equiv \frac{\mathrm{d}}{\mathrm{d}t}D(\rho_{1}(t),\rho_{2}(t))$ is also positive \cite{breuer2009measure}. These properties provide evidence that the proposed HSS-based witness is a bona-fide identifier of non-Markovianity.

\subsection{One-qubit systems}

\subsubsection{Phase-covariant noise}
We start by considering a single qubit undergoing a so-called phase covariant noise. The general time-local master equation, in the interaction picture (in units of $\hbar$), for the density matrix $\rho$ for a single qubit subject to phase-covariant noise is written as \cite{lankinen2016complete,smirnePRL,Teittinen_2018}
\begin{equation}\label{Mastercovaiant}
\dfrac{\mathrm{d}\rho}{\mathrm{d}t}=-i\omega (t)[\sigma_{z},\rho]+
\sum_{i=1}^3\frac{\gamma_{i}(t)}{2}L_{i}(\rho),
\end{equation}
where $ \omega(t) $ represents a time-dependent frequency shift, $ \gamma_{i}(t) $ ($i=1,2,3$) denotes the time-dependent rate associated to each dissipator $ L_{i}(\rho) $, whose expressions are \cite{lankinen2016complete}
\begin{eqnarray}
&&L_{1}(\rho)=\sigma_{+}\rho\sigma_{-}-\frac{1}{2}\{\sigma_{-}\sigma_{+},\rho\},\nonumber\\
&& L_{2}(\rho)=\sigma_{-}\rho\sigma_{+}-\frac{1}{2}\{\sigma_{+}\sigma_{-},\rho\},\nonumber \\
&& L_{3}(\rho)=\sigma_{z}\rho\sigma_{z}-\rho.
\end{eqnarray}
In the above equations, $ \sigma_{\pm} = \frac{1}{2}(\sigma_{x} \pm i\sigma_{y}) $ denote the inversion operators and $ \sigma_{i}$'s ($i = x,y,z$) are the Pauli operators. Moreover,  the three dissipators $L_{1},~ L_{2}$, and $ L_{3} $ describe, respectively, the heating, dissipation, and dephasing. Special cases of master equations of the form of Eq.~(\ref{Mastercovaiant}), describing the phase-covariant noise, are the amplitude damping model obtained for $ \gamma_{1}(t)=\gamma_{3}(t)=0 $ and the pure dephasing model achieved for $ \gamma_{1}(t)=\gamma_{2}(t)=0 $ \cite{breuer2016colloquium,nielsen00,he2019non}.

Indicating with $ |0\rangle $ and $ |1\rangle $ the ground and excited states
of the qubit, respectively, one can show that the solution of the master equation of Eq.~(\ref{Mastercovaiant}) is given by \cite{lankinen2016complete}
\begin{equation}\label{rhotPC}
\mathcal{E}_{t}(\rho(0))=\rho(t)=
\left( \begin {array}{cc} P_{1}(t)&Q(t)\\ \noalign{\medskip} Q^{*}(t)&1-P_{1}(t)\end {array} \right),
\end{equation}
where
\begin{equation}
P_{1}(t)=\text{e}^{-\Gamma(t)}[G(t)+P_{1}(0)],\
Q(t)=\alpha(0)\text{e}^{i\Omega(t)-\Gamma(t)/2-\tilde{\Gamma}(t)},
\end{equation}
with the time-dependent functions 
\begin{eqnarray}
&&\Gamma(t)=\int_{0}^{t}dt'[\gamma_{1}(t')+\gamma_{2}(t')]/2,\quad \tilde{\Gamma}(t)=\int_{0}^{t'}dt'\gamma_{3}(t'),\nonumber \\
&&\Omega(t)=\int_{0}^{t'}dt'2\omega(t'),\ 
G(t)=\int\limits_{0}^{t'}dt'\text{e}^{\Gamma(t')}\gamma_{2}(t')/2.
\end{eqnarray}
The master equation of Eq.~(\ref{Mastercovaiant}) leads to commutative dynamics, meaning $\mathcal{E}_{t}\circ\mathcal{E}_{s}=\mathcal{E}_{s}\circ\mathcal{E}_{t}  $ for any $ s $,
$ t\geq 0 $, iff $ \gamma_{1}(t)=\gamma(t) $ and $ \gamma_{2}(t)=\kappa \gamma(t) $, in which $0 \leq \kappa \leq 1 $. Moreover, the dynamics is unital, i.e. the corresponding channel $ \mathcal{E}_{t} $ satisfies $\mathcal{E}_{t}(\mathbb{I})=\mathbb{I}  $ ($ \mathbb{I} $ denotes the identity operator), when it is commutative and $\kappa =1 $.

Preparing the qubit in the initial state 
\begin{equation}\label{newpar}
|\psi_{0}\rangle=\dfrac{1}{\sqrt{2}}(\text{e}^{i\varphi}|+\rangle+|-\rangle),
\end{equation}
where $|\pm\rangle=\frac{1}{\sqrt{2}}(|0\rangle\pm|1\rangle)$, the time derivative of the HSS, that is the quantity $\chi(t)$ of Eq.~(\ref{chit}), results to be
\begin{eqnarray}
\chi(t)&=&
-\frac{1}{8}{\rm e}^{-2\tilde{\Gamma} (t) } \frac { \left( 
		\gamma_{1}(t)+\gamma_{2}(t)+4\gamma_{3}(t) \right) \cos^{2}\varphi }{\sqrt {{\rm e}^{\Gamma(t)-2\tilde{\Gamma} (t)} \cos^{2}\varphi  +
			\sin^{2}  \varphi  }}  \nonumber\\ 
&&-\frac{1}{4}{\rm e}^{-\Gamma (t) } \frac { \left( 
		\gamma_{1}(t)+\gamma_{2}(t)\right) \sin^{2}\varphi }{\sqrt {{\rm e}^{\Gamma(t)-2\tilde{\Gamma} (t)} \cos^{2}\varphi  +
			\sin^{2}  \varphi  }}	.
\end{eqnarray}
Accordingly, choosing $ \varphi = 0 $, the HSS-based witness $\chi(t)>0$ tells us that the process is non-Markovian when $\gamma_{1}(t)+\gamma_{2}(t)+4\gamma_{3}(t)<0$. On the other hand, choosing $ \varphi  = \frac{\pi}{2} $, the dynamics is non-Markovian by the HSS-based witness when $  \gamma_{1} (t) + \gamma_{2} (t)< 0 $. In other words, the dynamics is detected as non-Markovian if either of the conditions above holds. These conditions are a clear signature of dynamical information backflows, as can be deduced from the non-monotonicity of the off-diagonal terms of the evolved density matrix of Eq.~(\ref{rhotPC}). In fact, these conditions for $\chi(t)>0$ are exactly the ones that give $\sigma(t)>0$ (positive BLP witness) for the same dynamical instance \cite{teittinen2018revealing}): 
$\chi(t)>0 \Leftrightarrow \sigma(t) > 0$. Certainly, contractivity of the HSS is assured for Markovian conditions for which $\chi(t)<0$ for any $t>0$. 
Notice that the sensitivity of the witness $\chi(t)$ is investigated by considering general conditions for the phase-covariant noise, which encompass many of the most studied qubit dynamics such as pure dephasing, amplitude damping noise, depolarizing noise and the so-called eternal non-Markovianity \cite{hall2014canonical}. As a general insight from this first example, we thus observe that the HSS-based witness performs in perfect agreement with the BLP measure. It is known that the BLP measure, for which breaking CP-divisibility is a consequence of breaking P-divisibility \cite{breuer2009measure,PhysRevA.90.022110}, is tighter than other proposed non-Markovianity measures \cite{teittinen2018revealing}. On the basis of the above results, the same property holds for the HSS-based witness.

\subsubsection{Pauli channel}
In this section, we consider a qubit subject to a Pauli channel, whose corresponding master equation is \cite{chruscinski2013non,jiang2013comparing}
\begin{equation}
\dfrac{\mathrm{d}\rho}{\mathrm{d}t}=\sum_{i=1}^{3}\gamma_{i}(t)(\sigma_{i}\rho \sigma_{i}-\rho),
\end{equation}
where $ \gamma_{i}(t) $ ($i =1,2, 3$) denote the decoherence rate associated to the $i$-th channel. The dynamics may be rewritten in the following equivalent form \cite{chruscinski2013non,jiang2013comparing}
\begin{equation}
\rho(t)= \mathcal{E}_{t} [\rho(0)] =\sum\limits_{i=0}^{3}p_{i}(t)\sigma_{i}\rho(0)\sigma_{i},~~t\geq 0
\end{equation}
where $ \sigma_{0} = \mathbb{I} $ (identity operator), $ \sigma_{i}$'s are the Pauli matrices, and $ p_{i}(t) $'s denote the time-dependent probability distribution. Notice that $ p_{0}(0) = 1 $ and $p_i(0)=0$ ($i=1,2,3$), guaranteeing that $ \mathcal{E}_{0} = \mathbf{I} $ (identity channel). The explicit expressions of the time-dependent probabilities of the Pauli channel are
\begin{eqnarray}
&&p_{0}(t)=\dfrac{1}{4}[1+\lambda_{1}(t)+\lambda_{2}(t)+\lambda_{3}(t)],\nonumber \\
&& p_{1}(t)=\dfrac{1}{4}[1+\lambda_{1}(t)-\lambda_{2}(t)-\lambda_{3}(t)],\nonumber\\
&& p_{2}(t)=\dfrac{1}{4}[1+\lambda_{2}(t)-\lambda_{1}(t)-\lambda_{3}(t)],\nonumber\\
&&p_{3}(t)=\dfrac{1}{4}[1+\lambda_{3}(t)-\lambda_{2}(t)-\lambda_{1}(t)],
\end{eqnarray}
where
$\lambda_{1}(t)=\text{e}^{-2(\Gamma_{2}(t)+\Gamma_{3}(t))}$, 
$\lambda_{2}(t)=\text{e}^{-2(\Gamma_{1}(t)+\Gamma_{3}(t))}$, and
$\lambda_{3}(t)=\text{e}^{-2(\Gamma_{1}(t)+\Gamma_{2}(t))}$,
with 
\begin{equation}
\Gamma_{i}(t)=\int_{0}^{t}\gamma_{i}(\tau)\mathrm{d}\tau.\quad (i=1,2,3)
\end{equation}
It is straightforward to show that this dynamics is unital ($ \mathcal{E}_{t} (\mathbb{I}) = \mathbb{I}$). When $ \gamma_{1}{(t)} = \gamma_{2}{(t)} $, the unital case of the phase-covariant master equation and the Pauli channel with the same decay rates coincide with each other. It should be noted that the general Pauli channel includes a larger set of dynamics than the unital phase-covariant noise, such as bit-flip and bit-phase-flip channels. 

We now calculate the HSS-based witness $\chi(t)$ introduced in Eq.~(\ref{chit}), with the qubit initially prepared in a state parametrized as 
\begin{equation}
 |\psi_{0}^\pm(\varphi)\rangle=\frac{1}{\sqrt{2}}(\text{e}^{i\varphi}|0\rangle\pm |1\rangle).
\end{equation}
For three different optimal initial parametrizations given by the set $\{\ket{\psi_{0}^+(0)},\ket{\psi_{0}^+(\pi/2)},\ket{\psi_{0}^-(\pi/2)}\}$ one easily finds, respectively,  
\begin{eqnarray}
&&\chi(t)= - \left( \gamma_{1}(t)+\gamma_{3}(t) \right)  {{\rm e}^{-2{
					\Gamma_{1}(t)}-2{ \Gamma_{3}(t)}}} ,
					\nonumber \\
&& \chi(t)= - \left( \gamma_{1}(t)+\gamma_{2}(t) \right)   {{\rm e}^{-2{
					\Gamma_{1}(t)}-2{ \Gamma_{2}(t)}}} ,  
					\nonumber\\
&& \chi(t)=- \left( \gamma_{2}(t)+\gamma_{3}(t) \right)  {{\rm e}^{-2{
					\Gamma_{2}(t)}-2{ \Gamma_{3}(t)}}}. 
\end{eqnarray}
Therefore, according to the HSS-based criterion the dynamics is deemed Markovian if and only if 
$ \gamma_{1} (t) + \gamma_{2} (t)  \geq 0$, $\gamma_{1} (t) + \gamma_{3} (t) \geq 0$ and 
$\gamma_{2} (t) + \gamma_{3} (t) \geq 0 $ for all $ t\geq 0 $. Whenever at least one of the three conditions above is not satisfied, that is $ \gamma_{i}(t) + \gamma_{j}(t) < 0$ for some $ j\neq i $, one gets $\chi(t)>0$ so that the qubit dynamics exhibits memory effects and is non-Markovian. The latter is exactly the same condition that makes $\sigma(t)>0$ (positive BLP witness) \cite{chruscinski2013non,jiang2013comparing}, so once again: $\chi(t)>0 \Leftrightarrow \sigma(t) > 0$. In fact, it is well known that the qubit dynamics for the Pauli channel is Markovian according to BLP non-Markovianity criterion if and only if the sum of all pairs of distinct decoherence rates remains positive, i.e., $\gamma_{i}(t) + \gamma_{j}(t) \geq 0$ for all $ j\neq i $, for which contractivity of the HSS is verified ($\chi(t)<0$). Differently, for instance, according to the RHP non-Markovianity criterion, the dynamics is Markovian if and only if all of the decoherence rates remain positive for all $ t \geq 0  $, i.e., $ \gamma_{i}(t) \geq  0$, for all $i = 1,2,3$. Once again, the HSS-based witness is sensitive to system-environment information backflows in perfect agreement with the BLP measure.

\subsection{Two-qubit system}\label{Two-qubit example}
We now investigate a composite quantum system consisting of two separated qubits, A and B,  which independently interact with their own dissipative reservoir (leaky cavity). The general Hamiltonian is therefore written as $H=H_A+H_B$. The single qubit-reservoir Hamiltonian is ($\hbar\equiv 1$) \cite{breuer2002theory}
\begin{equation}
H=\omega_{0}~\sigma_{+}\sigma_{-}+\sum\limits_{k}^{}\omega_{k}b^{\dagger}_{k}b_{k}+(\sigma_{+}B+\sigma_{-}B^{\dagger}),
\end{equation}
where $ \omega_{0} $ represents the transition frequency of the qubit, $\sigma_{\pm}  $ are the system raising and lowering operators, $ \omega_{k} $ is the frequency of the $k$-th field mode of the reservoir, $ b_{k} $ and $ b^{\dagger}_{k} $ denote, respectively, the 
$k$-mode creation and annihilation operators, $ B=\sum_{k}^{}g_{k}b_{k} $ with $ g_{k} $ being the coupling constant with the $k$-th mode. At zero temperature and in the basis $ \{|1\rangle,|0\rangle\} $, from the above Hamiltonian with a Lorentzian spectral density for the cavity modes, one finds that the dynamics of the qubit can be described by the evolved reduced density matrix \cite{breuer2002theory,bellomo2007non}
\begin{equation}
\rho_\mathrm{q}(t)=\left( \begin {array}{cc} \rho_{11}^{S}(0)P(t)&\rho_{10}^{S}(0)\sqrt{P(t)}\\ \noalign{\medskip} \rho_{01}^{S}(0)\sqrt{P(t)}&1-\rho_{00}^{S}(0)P(t)\end {array} \right),
\end{equation}
where the coherence characteristic function $P(t)$ is
\begin{equation}\label{Pt}
P(t)=\text{e}^{-\lambda t}\left[\cos(\Gamma t/2)+(\lambda/\Gamma)\sin(\Gamma t/2)\right]^{2},
\end{equation}
with $ \Gamma=\sqrt{2\gamma_{0}\lambda-\lambda^{2}} $. The rate $ \lambda $ denotes the spectral width for the qubit-reservoir coupling (photon decay rate) and is connected to the reservoir correlation time $ \tau_{c} $ by the relation $ \tau_{c} =1/\lambda$. The decay rate $\gamma_{0}  $ is instead related to the system (qubit) relaxation time scale $\tau_{r}  $ by $ \tau_{r} =1/\gamma_{0}$. In the strong coupling regime, occurring for $ \gamma_{0} > \lambda/2 $, the non-Markovian effects become relevant \cite{breuer2002theory}. 

\begin{figure} [t!]
 \centering
	\includegraphics[width=0.52\textwidth]{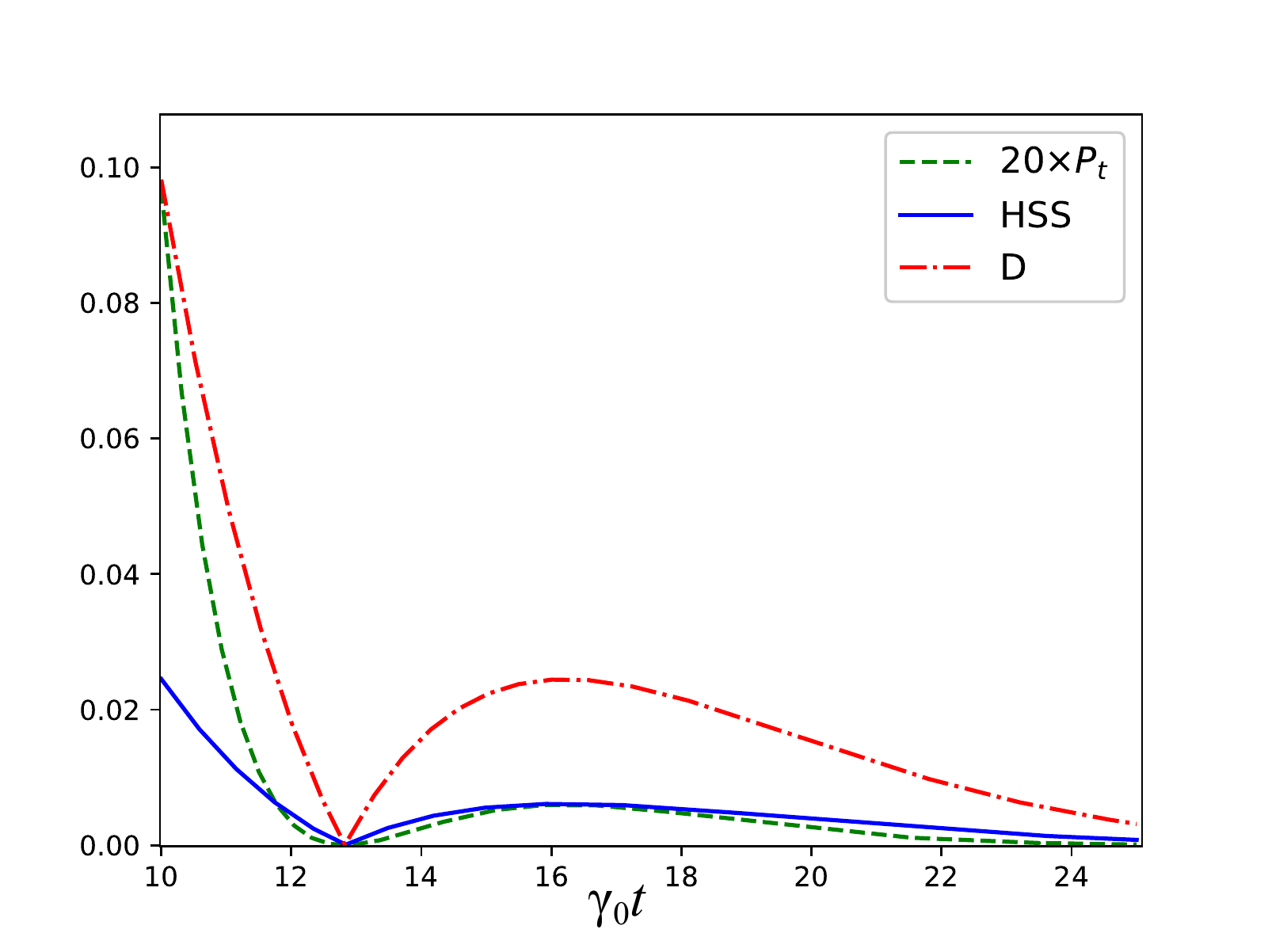}
	\caption{Dynamics of Hilbert-Schmidt speed $HSS(\rho_{\varphi}(t)) $ (blue solid line), trace distance $D(\rho_{1}(t),\rho_{2}(t))$ (red dot-dashed line) and coherence characteristic function $ P(t)$ (amplified by $20$ times for comparison, green dashed line) as a function of the dimensionless time $\gamma_0 t$ for the two-qubit system in the strong coupling regime, with $ \lambda=1.25 \gamma_0$. }\label{compare2}
\end{figure} 

The density matrix evolution of the two independent qubits can be then easily obtained knowing the evolved density matrix of a single qubit \cite{bellomo2007non}. The elements of the two-qubit evolved density matrix $ \rho(t) $ are presented in Appendix~\ref{A}.
Preparing the two-qubit system in the initial state 
\begin{equation}\label{initial2}
|\psi_{0}\rangle=\frac{1}{2}(\text{e}^{i\varphi}|11\rangle+|10\rangle+|01\rangle+|00\rangle),
\end{equation}
we find that the HSS of Eq.~(\ref{HSS}) is given by
\begin{equation}\label{HSS2}
HSS(\rho_{\varphi}(t))=\frac{1}{4}\sqrt {P(t) [ P(t) \left( 4\,P(t)-3 \right) +2 ] },
\end{equation}
which is independent of the phase $\varphi$. 
From this equation and from Eq.~(\ref{chit}), one promptly gets that the two-qubit dynamics is non-Markovian whenever
\begin{equation}
\chi(t)= \frac{\mathrm{d}HSS}{\mathrm{d}P}\frac{\mathrm{d}P}{\mathrm{d}t}>0,
\end{equation}
where $P=P(t)\in[0,1]$ is the coherence characteristic function of Eq.~(\ref{Pt}). Since 
$\mathrm{d}HSS/\mathrm{d}P$ is always positive, as easily seen from Eq.~(\ref{HSS2}), we obtain that $\chi(t)>0 \Leftrightarrow \mathrm{d}P/\mathrm{d}t>0$, a clear signature of information backflows from the environment to the system. So, this is also expected to happen for $\sigma(t)>0$ (positive BLP witness). 
Using the definition $D(\rho_{1},\rho_{2})=\frac{1}{2}\text{Tr}|\rho_{1}-\rho_{2}|$
and the optimal pair of two-qubit quantum states $\rho_{1}(0)=\ket{++}\bra{++}  $, $\rho_{2}(0)=\ket{--}\bra{--} $ with $ |\pm\rangle=\frac{1}{\sqrt{2}}(|0\rangle\pm|1\rangle) $, the time-dependent trace distance is \cite{wang2018probing}
\begin{equation}\label{TD2}
D(\rho_{1}(t),\rho_{2}(t))=\sqrt{P(t)(2-2P(t)+P(t)^{2})}.
\end{equation}
As a consequence, from $\sigma(t)=\frac{\mathrm{d}}{\mathrm{d}t}D(\rho_{1}(t),\rho_{2}(t))$, we have $\sigma(t)=\frac{\mathrm{d}D}{\mathrm{d}P}\frac{\mathrm{d}P}{\mathrm{d}t}$. Seeing that $\mathrm{d}D/\mathrm{d}P$ is always positive, one finds that $\sigma(t)>0$ whenever $\mathrm{d}P/\mathrm{d}t>0$, as expected. We hence obtain: $\chi(t)>0 \Leftrightarrow \sigma(t)>0$. 
The computation immediately shows that, in the weak coupling regime ($ \lambda>2\gamma_0 $), the behavior of $ D(\rho_{1}(t),\rho_{2}(t)) $, $HSS(\rho_{\varphi}(t))  $, and $ P(t) $ is essentially a Markovian exponential decay controlled by $ \gamma_{0} $ (all of them are decreasing monotonic functions of time): $\chi(t)$ and $\sigma(t)$ are always negative, verifying contractivity of the HSS. Differently, in the strong coupling regime ($ \lambda < 2\gamma_0 $), where memory effects arise, $ D(\rho_{1}(t),\rho_{2}(t)) $, $HSS(\rho_{\varphi}(t))  $, and $ P(t) $ simultaneously exhibit an oscillatory behavior such that their maximum and minimum points exactly coincide, as quantitatively shown in Fig.~\ref{compare2}. This two-qubit dissipative model also leads to the conclusion that the HSS-based witness of non-Markovianity is equivalent to the trace distance-based measure.

\subsection{One-qutrit systems}

\subsubsection{V-type three-level open quantum system}
In this section, we investigate the non-Markovian dynamics of a V-type three level atom, playing the role of a qutrit, coupled to a dissipative environment \cite{scully_zubairy_1997,gu2012non}. We recall that three-level quantum systems (qutrits) can be promising alternative candidates to be used in quantum processors instead of the standard two-level systems (qubits) \cite{lanyon2008,kumar}.
For a V-type qutrit interacting with a dissipative reservoir, the two upper levels, i.e., $ |2\rangle $ and $ |1\rangle $ are coupled to the ground state $ |0\rangle $ with transition frequencies $ \omega_{2} $ and $ \omega_{1} $, respectively. The Hamiltonian of the total system can be written as
\begin{equation}
H =H_{0} + H_{I},
\end{equation}
where ($\hbar\equiv 1$)
\begin{equation}
H_{0} =\sum\limits_{j=1}^{2}\omega_{j}\sigma^{(j)}_{+}\sigma^{(j)}_{-}+\sum\limits_{k}^{} \omega_{k} b^{\dagger}_{k}b_{k},
\end{equation}
represents the free Hamiltonian of the system plus the environment, while
\begin{equation}
H_{I} =\sum\limits_{j=1}^{2}\sum\limits_{k}^{}(g_{jk}\sigma^{(j)}_{+}b_{k}+g^{*}_{ik}\sigma^{(j)}_{-}b^{\dagger}_{k}),
\end{equation}
is the interaction Hamiltonian in which $\sigma^{(j)}_{\pm}$ ($j=1,2 $)
are the standard raising and lowering operators between each of the two upper levels and the ground one. The index $ k $ denotes the different reservoir field modes with frequencies $ \omega_{k} $, creation and annihilation operators $b^{\dagger}_{k}$, $b_{k} $ and coupling constants $ g_{jk} $.

We assume that  the relaxation rates of the two upper levels are equal, the two upper atomic levels are degenerated, the atomic transitions are resonant with the central frequency of the reservoir and the photonic bath is initially with no excitation. Under these conditions and after applying the unitary transformation 
\begin{equation}
\varrho(t)=U\rho_{S}(t)U^{\dagger},
\end{equation}
with
\begin{equation}
U=\left( \begin {array}{ccc} \dfrac{1}{\sqrt{2}}&-\dfrac{1}{\sqrt{2}}&0\\ \dfrac{1}{\sqrt{2}}&\dfrac{1}{\sqrt{2}}&0\\ \noalign{\medskip} 0&0&1\end {array} \right),
\end{equation}
on the evolved density matrix $ \rho(t) $ obtained in the interaction picture and written in the basis $ \{|2\rangle,~|1\rangle,~|0\rangle\} $,
one obtains the evolved state of the V-type atom by \cite{behzadi2017effect,gu2012non} 
\begin{equation}
\varrho(t)=\sum\limits_{i=1}^{3}\mathcal{K}_{i}\varrho(0)\mathcal{K}^{\dagger}_{i}.
\end{equation}
In the above dynamical map, the Kraus operators are
\begin{eqnarray}
&&\mathcal{K}_{1}=\left( \begin {array}{ccc} G_{+}(t)&0&0\\ 0&G_{-}(t)&0\\ \noalign{\medskip} 0&0&1\end {array} \right),\nonumber\\
&& \mathcal{K}_{2}=\left( \begin {array}{ccc} 0&0&0\\ 0&0&0\\ \noalign{\medskip} \sqrt{1-|G_{+}(t)|^{2}}&0&0\end {array} \right),\nonumber\\
&&\mathcal{K}_{3}=\left( \begin {array}{ccc} 0&0&0\\ 0&0&0\\ \noalign{\medskip} 0&\sqrt{1-|G_{-}(t)|^{2}}&0\end {array} \right),
\end{eqnarray}
with
\begin{equation}\label{Gpm}
G_{\pm}(t)=\text{e}^{-\lambda t/2}\bigg[\text{cosh}\bigg(\dfrac{d_{\pm} t}{2}\bigg)+\dfrac{\lambda}{d_{\pm}}\text{sinh}\bigg(\dfrac{d_{\pm} t}{2}\bigg)\bigg],
\end{equation}
where $d_{\pm}=\sqrt{\lambda^{2}-2\lambda \gamma (1\pm |\theta|)} $, $ \lambda $ is the spectral width of the reservoir, $ \gamma $ is the relaxation rate of the two upper levels to the ground state, and $ \theta $ depends on the relative angle between two dipole moment elements associated with the transitions $ |2\rangle \rightarrow |0\rangle $ and $ |1\rangle \rightarrow |0\rangle $. For example, $ \theta=0 $ means that the dipole moments of the two transitions are perpendicular to each other and corresponds to the case where there is no spontaneously generated interference (SGI) between the two decay channels; differently, $ \theta = \pm 1 $ indicates that the two dipole moments are parallel or antiparallel, corresponding to the strongest SGI between the two decay channels. Moreover, the two coherence characteristic functions $G_{\pm}(t)$ are associated, respectively, to the decay channels 
$\ket{\pm}\rightarrow\ket{0}$, where $\ket{\pm}=(\ket{2}\pm \ket{1})/\sqrt{2}$ \cite{behzadi2017effect,gu2012non}.

\begin{figure} [t!]
 \centering
	\includegraphics[width=0.52\textwidth]{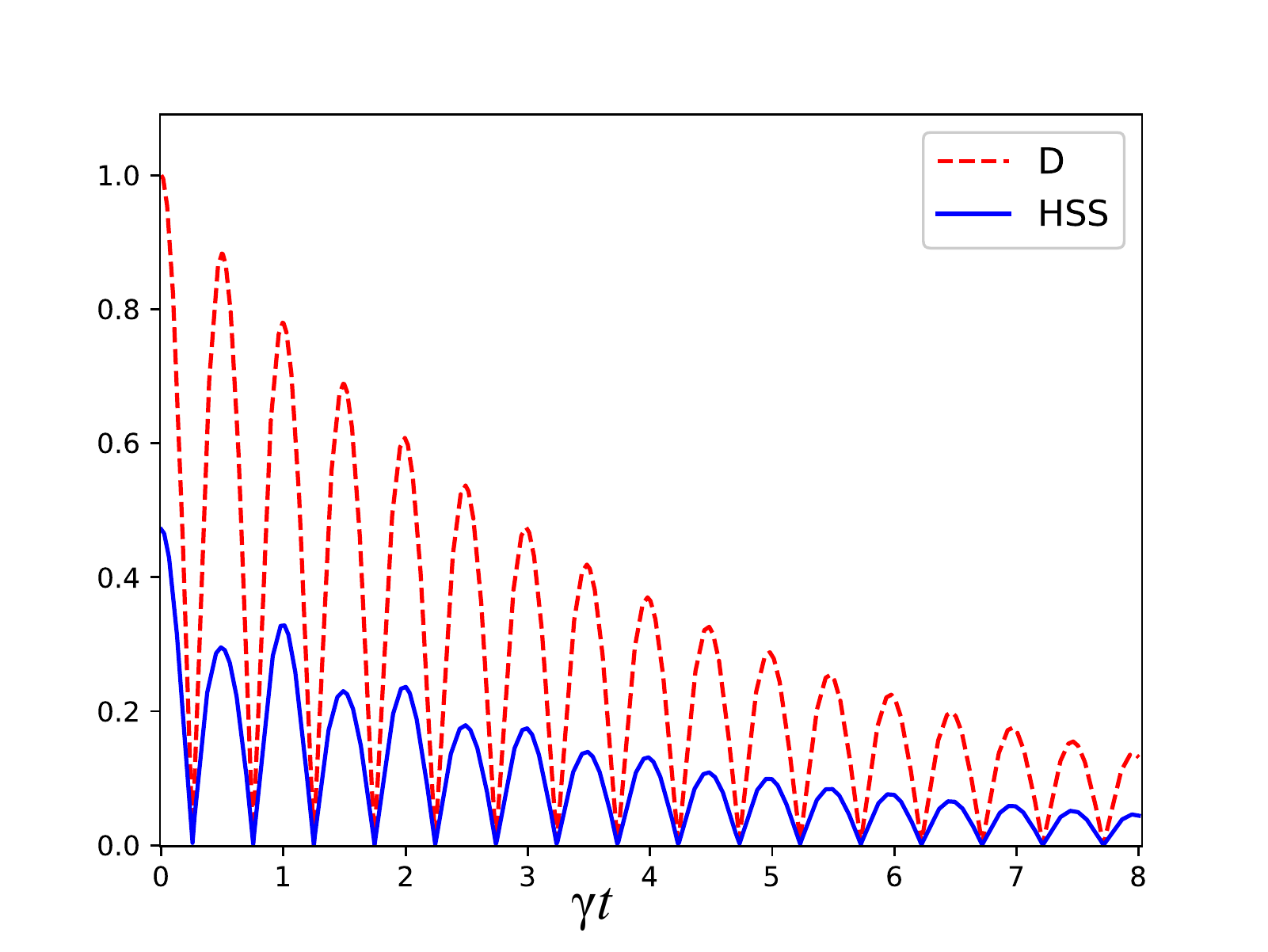}
	\caption{Dynamics of Hilbert-Schmidt speed $HSS(\rho_{\varphi}(t)) $ (blue solid line) and trace distance $D(\rho_{1}(t),\rho_{2}(t))$ (red dashed line) as a function of the dimensionless time $\gamma t$ for the V-type three-level atom, with $\lambda=5\times10^{-3}\gamma $ and $ \theta=0.6 $.}\label{VTYPEHSS}
\end{figure} 

To assess the memory effects by the HSS-based measure, the qutrit is initially taken in the state 
\begin{equation}
|\psi_{0}\rangle=\dfrac{1}{\sqrt{3}}(\text{e}^{i\varphi}|\tilde{2}\rangle+|\tilde{1}\rangle+|\tilde{0}\rangle), 
\end{equation}
where $|\tilde{i}\rangle  =U|i\rangle$ ($i=1,2,3$). The HSS of Eq.~(\ref{HSS}) is then easily obtained as 
\begin{equation}\label{HSSV}
HSS(\varrho_{\varphi}(t))=\frac{1}{3}|G_{+}(t)|\sqrt{|G_{-}(t)|^{2}+1},
\end{equation}
being independent of the initial phase $\varphi$.
Firstly we notice that, as physically expected, the HSS above depends on both $G_{\pm}(t)$ so taking into account the interplay (interference effects) of the two decay channels. Also, under Markovian (memoryless) evolution of the qutrit, occurring for $\lambda>4\gamma$ (weak-coupling regime), $HSS(\varrho_{\varphi}(t))$ is monotonically decreasing and thus contractive. Memory effects are therefore detected when $\chi(t)>0$, that is when the combination of the two channel contributions provides a net information backflow from the environment to the system (quantum speedup).
On the other hand, the trace distance-based measure, obtained by choosing a pair of initial orthogonal pure states $ \rho_{1}(0)=|\psi_{+}\rangle \langle \psi_{+}|$ and $\rho_{2}(0)=|\psi_{-}\rangle \langle \psi_{-}| $, where $ |\psi_\pm\rangle=(|+\rangle\pm |0\rangle)/\sqrt{2}$, is given by \cite{gu2012non}
\begin{equation}\label{DV}
D(\rho_{1}(t),\rho_{2}(t))=|G_{+}(t)|.
\end{equation}
It is worth to notice that this expression does not encompass the contribution due to $|G_{-}(t)|$ governing the decay channel $\ket{-}\rightarrow\ket{0}$, which makes us doubt whether the pair of initial states $ \rho_{1}(0)$, $ \rho_{2}(0)$ above is really the optimal one or not. Indeed, it is known that maximizing the trace distance for systems with dimension larger than 2 may be a challenging task in general. However, for $\theta = 0, \pm1$, from Eqs.~(\ref{Gpm}), (\ref{HSSV}) and (\ref{DV}), one immediately finds that the qualitative dynamics of $HSS(\varrho_{\varphi}(t))$ and $D(\rho_{1}(t),\rho_{2}(t))$ perfectly agree, giving: $\chi(t)>0\Leftrightarrow \sigma(t)>0$. For intermediate values of the parameter $\theta$, $HSS(\varrho_{\varphi}(t))$ and $D(\rho_{1}(t),\rho_{2}(t))$ maintain the general property of having the same zeros (in the oscillatory strong-coupling regime, $\lambda<4\gamma$), but their maximum points do not exactly coincide (we recall that this may be due to a nonoptimal choice of the initial states for maximizing the trace distance). 
The more intensely the strong coupling regime is satisfied ($\lambda\ll \gamma$, that means stronger memory effects), the tighter the accordance between their maximum points appears. 
The time behaviors of the two non-Markovianity witnesses are plotted in Fig.~\ref{VTYPEHSS} for $\theta=0.6$ and $\lambda = 5\times10^{-3}\gamma$. One can appreciate that the trace distance and the HSS exhibit an excellent qualitative agreement, with very close maximum points. Overall, the HSS-based measure results to be a valid non-Markovianity identifier for this open V-type qutrit dynamics.

\subsubsection{$\Lambda $-type three-level open quantum system}
The last system considered in our case study analysis is the so-called $\Lambda $ model, consisting of a three-level atom (qutrit) with excited state $ |a\rangle $ and two ground states $ |b\rangle $ and $ |c\rangle $ which interacts off-resonantly with a cavity field \cite{scully_zubairy_1997}. The cavity modes are assumed to have a Lorentzian spectral density 
\begin{equation}
J(\omega)=\dfrac{\gamma_{0}}{2\pi}\dfrac{\lambda^{2}}{(\omega_\mathrm{cav}-\omega)^{2}+\lambda^{2}},
\end{equation}
where $ \lambda $ is the cavity spectral width,  $ \omega_\mathrm{cav} $ represents the resonance frequency of the cavity, and the rate $\gamma_{0}  $ quantifies the strength of the system-environment coupling. Moreover, $ \Delta_{i}=\omega_{i}-\omega_{cav} $ denotes the detuning of the $ i $-th transition frequency of the atom from the cavity resonance frequency, being $\omega_1\equiv\omega_{ab}$ and $\omega_2\equiv\omega_{ac}$. The master equation describing the reduced dynamics of the $ \Lambda $-type atom and its analytical solution are reported, for convenience, in Appendix~\ref{B}. This is characterized by two Lindblad operators $\ket{b}\bra{a}$ and $\ket{c}\bra{a}$ corresponding to the time-dependent decay rates, respectively, $\gamma_1(t)$ and $\gamma_2(t)$. 

To find the conditions for dynamical memory effects by means of the HSS-based measure, we prepare the $\Lambda $-type atom in the initial state 
\begin{equation}
|\psi_{0}\rangle=\frac{1}{\sqrt{3}}(\text{e}^{i\varphi}|a\rangle+|b\rangle+|c\rangle), 
\end{equation}
which gives, from Eq.~(\ref{HSS}), $HSS(\rho_{\varphi}(t))=\frac{\sqrt{2}}{3}\text{e}^{-[\Gamma_{1}(t)+\Gamma_{2}(t)]/2} $, where $\Gamma_i(t)=\int_{0}^{t}\mathrm{d}s \gamma_{i}(s)$. 
Therefore, the non-Markovianity witness $\chi(t)$ of Eq.~(\ref{chit}) is
\begin{equation}
\chi(t)=\dfrac{-(\gamma_{1}(t)+\gamma_{2}(t))}{3\sqrt{2}}\text{e}^{-[\Gamma_{1}(t)+\Gamma_{2}(t)]/2}.
\end{equation}
This equation reveals that the non-Markovian character of the system dynamics is identified by the sum of the time-dependent decay rates $\gamma_{1}(t)+\gamma_{2}(t)$, which takes into account the competing processes of the two decay channels associated to $\gamma_{1}(t)$ and $\gamma_{2}(t)$, respectively. This is physically expected, also on the basis of previous analysis of such a $\Lambda$-type system in terms of non-Markovian quantum jumps \cite{piilo2009open}. 

Let us qualitatively discuss some particular conditions. 
As promptly seen from the canonical master equation given in Appendix~\ref{B}, if both the decay rates $\gamma_{1}(t)$, $\gamma_{2}(t)$ are nonnegative during the evolution, the open dynamics is Markovian (memoryless) \cite{hall2014canonical}, giving $\chi(t)\leq 0$ and so verifying contractivity of the HSS: in this case, the rate of information flow may change but the direction of the flow remains constant, namely from the system to the environment. 
On the other hand, it is known that, when the detunings $ \Delta_{i}$ are large enough, the decay rates $\gamma_i$ assume temporary negative values which produce information backflows from the cavity to the system \cite{piilo2009open,laine2010measure}: hence, memory effects occur ($\chi(t)>0$) when $\gamma_{1}(t)+\gamma_{2}(t)<0$ with an overall backflow of information. For $ \Delta_{1}= \Delta_{2}$ the decay rates are simultaneously negative in the same time regions, while for $ \Delta_{1} \neq \Delta_{2}$ the decay rates can have opposite signs \cite{laine2010measure}. In the latter situation, the cooperative action of the two channels become relevant. 
When the channel corresponding to the decay rate $\gamma_i (t)$ ($i = 1,2$) produces more information flow from environment to system than the other channel associated to $\gamma_j (t)$ ($j \neq i$), then $|\gamma_i (t)| > |\gamma_j(t)|$. This means that $\gamma_j (t) < -\gamma_i (t)$ during the time intervals when $\gamma_i(t)$ is negative and $\gamma_j(t)$ is positive: it is thus sufficient that only $\gamma_i(t)$ is negative to assure non-Markovianity ($\chi(t)>0$). 
These results are fully consistent with the previous findings obtained by the BLP (trace distace-based) witness and $\sigma(t)$ \cite{laine2010measure}. This open $\Lambda$-type qutrit system thus gives: $\chi(t)>0\Leftrightarrow \sigma(t)>0$, confirming the faithfulness of the HSS-based measure to detect memory effects in open quantum systems of dimension three.

\section{Conclusions}\label{cunclusion}

We have established a relation between the non-Markovian dynamics of open quantum systems and the positive changing rate of the Hilbert-Schmidt speed (HSS), which is a special case of quantum statistical speed. The idea underlying this definition is grounded on the fact that the nonmonotonic speed (positive acceleration) of quantum evolutions is a signature of memory effects in the dynamics of the system interacting with the surrounding environment. By the introduced HSS-based witness, one can then define a quantitative measure of dynamical memory effects. 

We have shown, in an extensive case study analysis, that the proposed witness is as efficient as the well-known trace distance-based (BLP) witness in detecting the non-Markovianity. 
The models considered for our study encompass many of the most paradigmatic open quantum systems (single qubits, two qubits and single qutrits undergoing dissipative and nondissipative dynamics), and supply evidence for the sensitivity of our HSS-based witness to system-environment information backflows. 
Besides its conceptual interest, we remark that the HSS-based witness does not require diagonalization of the reduced system density matrix, with consequent practical advantages in the analysis. In fact, a valid quantifier with this characteristic would be highly desired, especially for assessing memory effects of high-dimensional and multipartite open quantum systems. 

The HSS is related to the Hilbert-Schmidt metric. However, despite the noncontractivity of the Hilbert-Schmidt distance for quantum systems of dimension $n>2$, we have shown that the HSS-based witness is a faithful non-Markovianity measure (satisfying contractivity) for all the systems studied, including qutrits ($n=3$). As a prospect, these results stimulate the investigation for systems of higher dimension to assess the extent of validity.

Our study supplies an alternative useful tool to detect non-Markovianity based on the concept of quantum statistical speed detecting system-environment backflows of information. It thus motivates further analyses on the role of memory effects in composite open quantum systems and their relation to quantum speedup.

\section*{Acknowledgements}
H.R.J. thanks Henri Lyyra and Jose Teittinen for invaluable comments as well as constructive remarks and highly appreciates Sabrina Maniscalco for useful help. K.M. and M.K.S. would like to thank Farzam Nosrati for useful discussions. H.R.J. also wishes to acknowledge the financial support of the MSRT of Iran and Jahrom University.

\appendix
\section{Two-qubit evolved density matrix}\label{A}

Following the procedure described in Ref.~\cite{bellomo2007non} to construct the reduced density matrix $\rho(t) $ for the two-qubit system discussed in Sec.~\ref{Two-qubit example}, one finds that the diagonal and nondiagonal elements of $\rho(t)$ in the computational basis $  \{|11\rangle,|10\rangle,|01\rangle,|00\rangle\}$ are given by
\begin{eqnarray}
&&\rho_{11}(t)=\rho_{11}(0)P(t)^{2},\nonumber\\
&& \rho_{22}(t)=\rho_{22}(0)P(t)+\rho_{11}(0)P(t)(1-P(t)),\nonumber\\
&& \rho_{33}(t)=\rho_{33}(0)P(t)+\rho_{11}(0)P(t)(1-P(t)),\nonumber\\
&&\rho_{44}(t)=1-[\rho_{11}(t)+\rho_{22}(t)+\rho_{33}(t)],
\end{eqnarray}
and 
\begin{eqnarray}
&&\rho_{12}(t)=\rho_{12}(0)P(t)^{3/2},~~\rho_{13}(t)=\rho_{13}(0)P(t)^{3/2},\nonumber\\
&& \rho_{14}(t)=\rho_{12}(0)P(t),~~\rho_{23}(t)=\rho_{23}(0)P(t),\nonumber\\
&& \rho_{24}(t)=\sqrt{P(t)}[\rho_{24}(0)+\rho_{13}(0)(1-P(t))],\nonumber\\
&& \rho_{34}(t)=\sqrt{P(t)}[\rho_{34}(0)+\rho_{12}(0)(1-P(t))],
\end{eqnarray}
with $\rho_{ji}(t) =\rho^{*}_{ij}(t) $.

\section{Solutions for $\Lambda $-type three-level system}\label{B}

This appendix presents the formal analytical solutions for the $\Lambda $-type three-level systems \cite{piilo2009open,laine2010measure}. 
The weak-coupling master equation for this model is written as follows
\begin{align}
\frac{\mathrm{d}}{\mathrm{d}t}\rho(t)&=-i\lambda_{1}(t)[|a\rangle\langle a|,\rho(t)]-i\lambda_{2}(t)[|a\rangle\langle a|,\rho(t)]\nonumber\\
&+\gamma_{1}(t)\bigg[|b\rangle\langle a| \rho(t) |a\rangle\langle b|-\frac{1}{2}\{\rho(t),|a\rangle\langle a|\}\bigg]\nonumber\\
&+\gamma_{2}(t)\bigg[|c\rangle\langle a| \rho(t) |a\rangle\langle c|-\frac{1}{2}\{\rho(t),|a\rangle\langle a|\}\bigg],
\end{align}
where
\begin{eqnarray}
&&\lambda_{i}(t)=\int\limits_{0}^{t}\mathrm{d}s \int\limits_{0}^{\infty}\mathrm{d}s J(\omega) \text{sin}[(\omega-\omega_{i})s],\nonumber\\
&&\gamma_{i}(t)=\int\limits_{0}^{t}\mathrm{d}s \int\limits_{0}^{\infty}\mathrm{d}s J(\omega) \text{cos}[(\omega-\omega_{i})s].
\end{eqnarray}
Introducing the short-hand notation
\begin{equation}\label{DLi}
D_{i}(t)=\int_{0}^{t}\mathrm{d}s \gamma_{i}(s),\quad
L_{i}(t)=\int_{0}^{t}\mathrm{d}s\lambda_{i}(s),
\end{equation}
one finds that the solution of the master equation is given by \cite{piilo2009open,laine2010measure}
\begin{eqnarray}\label{gammadensity}
&&\rho_{aa}(t)=\rho_{aa}(0)\text{e}^{-[D_{1}(t)+D_{2}(t)]},\nonumber\\
&& \rho_{bb}(t)=\rho_{aa}(0)\int_{0}^{t}\mathrm{d}s \gamma_{1}(s)\text{e}^{-[D_{1}(s)+D_{2}(s)]}+\rho_{bb}(0),\nonumber\\
&& \rho_{cc}(t)=\rho_{aa}(0)\int_{0}^{t}\mathrm{d}s \gamma_{2}(s)\text{e}^{-[D_{1}(s)+D_{2}(s)]}+\rho_{cc}(0),\\
&& \rho_{ab}(t)=\rho_{ab}(0)\text{e}^{-[D_{1}(t)+D_{2}(t)]/2}\text{e}^{-i[L_{1}(t)+L_{2}(t)]},\nonumber\\
&& \rho_{ac}(t)=\rho_{ac}(0)\text{e}^{-[D_{1}(t)+D_{2}(t)]/2}\text{e}^{-i[L_{1}(t)+L_{2}(t)]},\nonumber\\
&& \rho_{bc}(t)=\rho_{bc}(0)\nonumber.
\end{eqnarray}


\end{document}